# Detection of bosonic mode as a signature of magnetic excitation in one-unit-cell FeSe on SrTiO$_3$

Chaofei Liu[1], Ziqiao Wang[1], Shusen Ye[1], Cheng Chen[1], Yi Liu[1], Qingyan Wang[1], Qiang-Hua Wang[2,3] and Jian Wang[1,4,5,†]

[1]*International Center for Quantum Materials, School of Physics, Peking University, Beijing 100871, China*
[2]*National Laboratory of Solid State Microstructures & School of Physics, Nanjing University, Nanjing 210093, China*
[3]*Collaborative Innovation Center of Advanced Microstructures, Nanjing University, Nanjing 210093, China*
[4]*Collaborative Innovation Center of Quantum Matter, Beijing 100871, China*
[5]*CAS Center for Excellence in Topological Quantum Computation, University of Chinese Academy of Sciences, Beijing 100190, China*

We report an *in situ* scanning tunneling spectroscopy study of one-unit-cell (1-UC) FeSe film on SrTiO$_3$(001) (STO) substrate. In quasiparticle density of states, bosonic excitation mode characterized by the "dip-hump" structure is detected outside the larger superconducting gap with energy comparable with phonon and spin resonance modes in heavily electron-doped iron selenides. Statistically, the excitation mode, which is intimately correlated with superconductivity, shows an anticorrelation with pairing strength and yields an energy scale upper-bounded by twice the superconducting gap coinciding with the characteristics of magnetic resonance in cuprates and iron-based superconductors. The local response of tunneling spectra to magnetically different Se defects all exhibits the induced in-gap quasiparticle bound states, indicating an unconventional sign-reversing pairing. These results support the magnetic nature of the excitation mode and possibly reveal a signature of electron–magnetic-excitation coupling in high-temperature superconductivity of 1-UC FeSe/STO.

The electron-phonon coupling in Bardeen-Cooper-Schrieffer (BCS) theory was unequivocally verified by tunneling experiments in conventional superconductors [1–4]. As the pairing "glue" of condensate Cooper pair, the bosonic excitation mode (phonon) is fingerprinted as a "dip-hump" structure in the quasiparticle–density-of-state (DOS) spectrum [4]. Generalized to cuprates and iron-based superconductors, the hump anomaly was analogously detected [5–12], resembling the phonon signature of conventional superconductors. In hole-doped Bi$_2$Sr$_2$CaCu$_2$O$_{8+\delta}$, the bosonic feature was ascribed to a phonon mode mainly based on the isotope-substitution–dependent behavior [6]. In contrast, some other tunneling experiments identified a spin-excitation origin of the bosonic mode by its doping dependence in widely ranged hole-doped Bi$_2$Sr$_2$CaCu$_2$O$_{8+\delta}$ [5] and Pr$_{0.88}$LaCe$_{0.12}$CuO$_4$ [7]. As for iron pnictides, the bosonic excitation was captured with energy consistent with a spin resonance mode observed by inelastic neutron scattering, hinting plausible spin-fluctuation–mediated superconductivity [9–12]. Although the interpretation is still under debate over phonon or magnetic excitation, the intensively studied bosonic signal has been validated to be essential in unveiling the enigmatic pairing symmetry of high-temperature superconductivity.

However, systematic investigations on the bosonic excitation mode remain scarce [13] in the recently discovered one-unit-cell (1-UC) FeSe on SrTiO$_3$(001) (STO) [14]. The 1-UC FeSe/STO grown by molecular-beam epitaxy (MBE), which ranks the highest critical temperature ($T_c$) (typically 65 K) in iron-based superconductors, has triggered tremendous research interest in condensed-matter physics [15–23]. Photoemission spectroscopy suggests that the Fermi surface consists of only electron pockets at M points [15], different from that found in bulk FeSe and iron pnictides with additional hole pockets at Γ point. In consequence, the previously proposed $s_\pm$-wave pairing, based on repulsive pair scatterings between electron and hole pockets with opposite signs of superconducting order parameters [24,25], is drastically challenged. Given the fine structure of two crossed elliptic electron pockets at M points in 1-UC FeSe [26], various pairing symmetries on the basis of such a Fermi-surface topology without hole pockets were theoretically proposed, e.g., sign-preserving $s_{++}$-wave [27,28], sign-reversing extended $s_\pm$-wave [29,30] and quasinodeless $d$-wave pairings [31–33]. Experimentally, the gate-tuned $T_c$ [34] and the STO-phonon–induced "replica" band [35] highlight the importance of charge doping and electron-phonon interaction (EPI), respectively. More directly, the similar energy evolution of intra- and inter-pocket scatterings shows evidence for electron-phonon–coupled $s$-wave pairing [36]. On the other hand, the gapped boson spectrum within excitonic resonance energy and the absence of STO phonon at grain-boundary–free regions speak against interfacial phonon coupling as the origin of colossal $T_c$ enhancement [37]. Furthermore, another heavily electron-doped iron selenide, $(Li_{1-x}Fe_x)OHFeSe$, which is in remarkable similarity with interfacial FeSe/STO in band structure, Fermi surface and gap symmetry [38,39], was quantitatively analyzed favoring a sign-reversal scenario [40]. Therefore, the underlying pairing mechanism of 1-UC FeSe still remains elusive.

In this Letter, we present a systematic investigation on the bosonic excitation mode in 1-UC FeSe/STO by *in situ* scanning tunneling microscopy/spectroscopy (STM/STS), aimed at gaining a deeper insight into the recently debated pairing symmetry therein. Our experiments were conducted in an ultrahigh-vacuum MBE-STM combined system (Scienta Omicron) (Part I in Ref. [41]). All the topographic images and tunneling spectra ($dI/dV$ vs. $V$) were measured at 4.2 K unless specifically emphasized. Figure 1(a) shows typical large-scale and atomically resolved topographic images of 1-UC FeSe film, indicative of a highly crystalline sample quality. An averaged tunneling spectrum is presented in Fig. 1(b). Double superconducting gaps ($\Delta_i$, $i$=1, 2) and the "U"-shaped flat bottom around zero bias can be seen, which are more clearly revealed in the normalized [Fig. 1(c)] and derivative ($d^2I/dV^2$ vs. $V$) [Fig. 1(d)] versions. A detailed statistics of $\Delta_1$ and $\Delta_2$ yields energies of 11.5 and 18.3 meV [Fig. 2(c)], respectively, comparable with previous reports [14,34,36].

An interesting feature [13] is rescrutinized: outside $\Delta_2$, there emerges a dip-hump structure [Fig. 1(c)], which is intrinsic and robust against averaging procedure. In strongly coupled conventional superconductors [3], cuprates [5–8] and iron-based superconductors [9–12,42], the dip-hump feature was analyzed as a signature of quasiparticle coupling to a collective boson spectrum [43]. With an extrapolated $T_c$ of 56.6 K (Fig. S3), our FeSe film corresponds to $2\Delta_i/k_BT_c$ ($k_B$, Boltzmann constant) of 4.7 and 7.5 for $\Delta_1$ and $\Delta_2$, respectively, indicating a strong-coupling nature. We thus identified the hump feature as a bosonic excitation mode, with energy $\Omega$ defined as the hump energy $\Delta_2+\Omega$ offset by $\Delta_2$. Along a straight-line trajectory on FeSe surface, the normalized tunneling spectra exhibit both superconducting gaps and excitation modes spatially [Fig. 1(e)], hinting a potential relevance of the bosonic mode to superconductivity.

To further confirm the relationship between the excitation mode and superconductivity, we studied intrinsic-defect and temperature dependence of the tunneling spectra. Figures 2(a) and 2(b) present the spectrum evolution towards a grain boundary and elevated temperatures, respectively. When superconducting order is completely destroyed at the boundary site, the bosonic feature disappears. Meanwhile, as increasing temperature towards $T_c$, the coherence peaks and the excitation mode progressively fade out. For comparison, the normalized 4.2-K spectrum is convol-

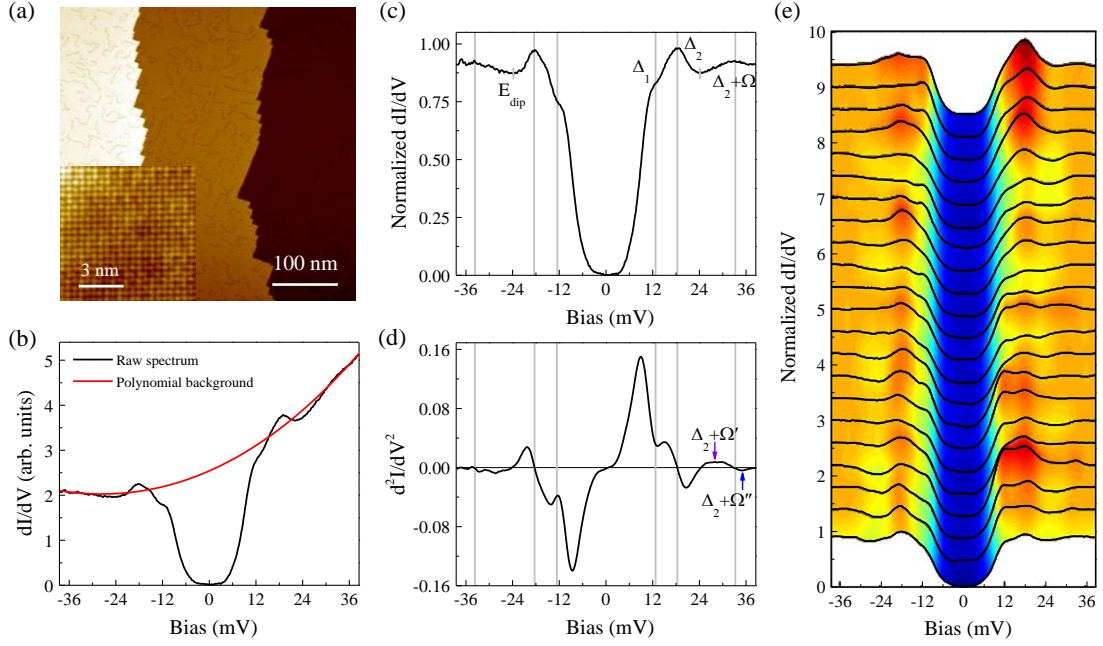

FIG. 1 (color online). (a) Topographic images of 1-UC FeSe film in different scales. (b) Averaged tunneling spectrum (black) over ~100 spectra and its polynomial background (red), which is extracted by a cubic-polynomial fitting for bias $|V| \geq 30$ mV. (c) Individually normalized tunneling spectrum by its own polynomial background (by default). $\Delta_1$, $\Delta_2$, $E_{dip}$ and $\Delta_2+\Omega$, which denote the superconducting gaps, the dip and hump energies, are assigned as half the energy separations between corresponding coherence peaks, dips or humps (gray). (d) Derivative of the normalized spectrum in (c). The most outer gray line, violet and blue arrows denote the zero-point ($\Delta_2+\Omega$), peak ($\Delta_2+\Omega'$) and dip positions ($\Delta_2+\Omega''$) (e.g., positive-bias side) of the derivative, respectively. The latter two define the hump energies as well in some literatures [6,7,10–12,42]. (e) Spatially resolved normalized tunneling spectra along a straight-line trajectory on FeSe surface.

uted by Fermi-Dirac distribution function at 6–35 K [solid curves in Fig. 2(b)]. Prominently, in the convoluted spectra, the progressive weakening of superconducting and bosonic signals is reproduced. All these phenomena highlight an intimate correlation between the bosonic excitation mode and superconductivity. In previous studies following Eliashberg strong-coupling theory, the bosonic mode was ascribed to phonon or spin fluctuation by comparing electron-boson spectral function $\alpha^2F(\omega)$ with DOS derivative spectrum [3,43]. Hence, identity of the detected bosonic mode should be studied in detail. We present the statistics of $\Delta_1$, $\Delta_2$ and the hump energy $\Delta_2+\Omega$ in Fig. 2(c), yielding an energy range of 9.6–18.2 meV for the bosonic mode (for two other definitions, $\Omega'$ and $\Omega''$, see Fig. S4). As shown later, the bosonic-mode energy $\Omega$ is dependent of annealing process and statistically anticorrelated with $\Delta_i$, which will be increased as decreasing $\Delta_i$. Specifically, the larger $\Delta_i$ of 1-UC FeSe film (11.5 and 18.3 meV) than those of $(Li_{0.8}Fe_{0.2})OHFeSe$ (9 and 15 meV [38]) will conclude an energy-diminished bosonic mode. Thereby, the smaller bosonic-energy range obtained here is reasonably comparable with a spin resonance mode of 21 meV observed by inelastic neutron scattering in $(Li_{0.8}Fe_{0.2})OHFeSe$ [44]. Since the 1-UC FeSe is too thin to be prevented from being blurred by unexpected signal of far thicker protection layer or substrate, it is challenging to carry out *ex situ* neutron-scattering experiments. Alternatively, it is acceptable to take the spin resonance mode in $(Li_{0.8}Fe_{0.2})OHFeSe$ as a reference for interfacial FeSe/STO in view of the above-mentioned similarities between them

[38,39].

We now discuss the phonon possibility of the bosonic mode. A 20-meV optical phonon detected by high-resolution electron-energy loss spectroscopy in 1-UC FeSe [45] is also noticed. Generally, the phonon signature tends to persist across $T_c$ [12], which is inconsistent with the fading-out behavior of the bosonic feature as increasing temperature in Fig. 2(b). To elucidate the underlying excitation nature in depth, we lengthened the annealing time and intentionally dosed magnetically distinct Fe and Pb adatoms upon the FeSe surface. Figure 2(d) shows the annealing dependence (1st–3rd spectra) and response to adatoms (4th–7th spectra) of averaged normalized tunneling spectra. All the spectra have been shifted leftwards by $\Delta_2$ to reveal the bosonic-mode energies $\Omega$ (arrows) directly, which are more evidently distinguished as zero points of the derivative spectra in Fig. S5. The unchanged bosonic-mode energies in different adatom-free regions (1st, 2nd, 4th and 6th spectra) verify the reliability of averaging and $\Omega$-determining procedures. It is found that $\Omega$ is annealing-dependent (1st/2nd vs. 3rd spectrum), i.e, sensitive to electron doping. As for adatom dosing, $\Omega$ responds more sensitively to Fe adatoms than to Pb (4th/5th vs. 6th/7th spectra). Therefore

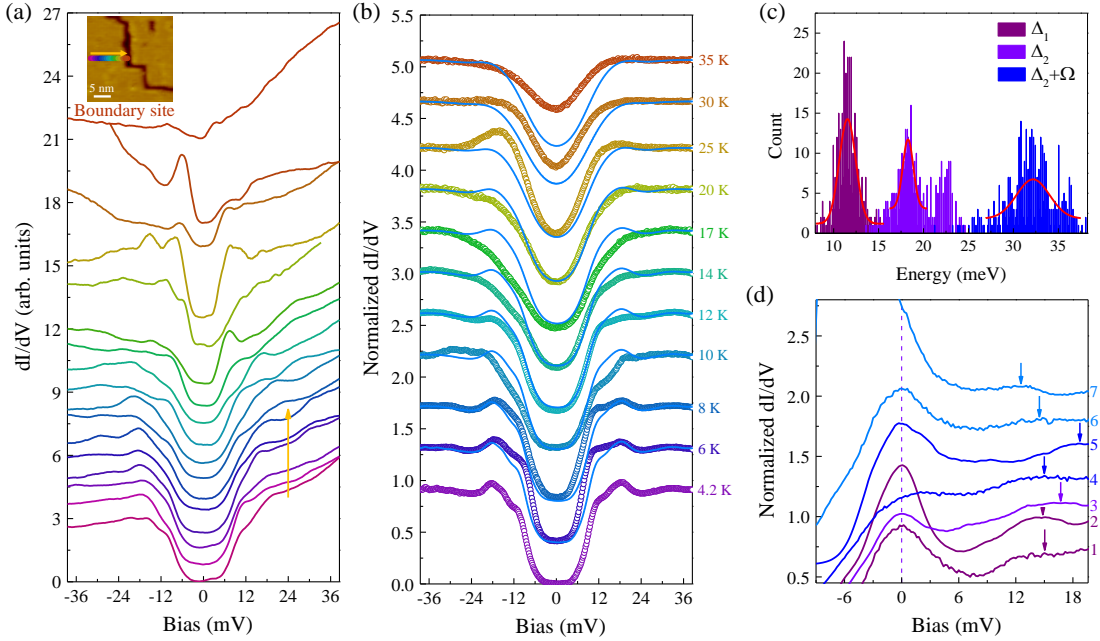

FIG. 2 (color online). (a) Spatial dependence of the tunneling spectra along a trajectory gradually approaching a grain boundary (inset; for details, see Fig. S1). (b) Temperature dependence of the normalized tunneling spectra (open symbols). The solid curves are convoluted normalized 4.2-K spectra by Fermi-Dirac distribution function. (c) Statistics of $\Delta_1$, $\Delta_2$ and the hump energy $\Delta_2+\Omega$ for 1-UC FeSe film annealed at 450 ℃ for 3 h (by default unless otherwise mentioned), all yielding Gaussian distributions (red) with expectation values of 11.5, 18.3 and 32.2 meV, respectively. The Gaussian fitting of $\Delta_2+\Omega$ statistics determines an energy range of 9.6–18.2 meV for the bosonic-mode energy $\Omega$. (d) Annealing dependence and response to adatoms of averaged (typically over ~100–200 spectra) normalized tunneling spectra. All the spectra have been rescaled for a better view of humps and shifted leftwards to center $\Delta_2$ at zero bias for a direct comparison of $\Omega$ (arrows). The 1st/2nd spectra: both annealed at 450 ℃ for 3 h but imaged at different regions ($\Omega$, 15/14.8 meV); 3rd: annealed at 450 ℃ for 10.7 h ($\Omega$, 16.7 meV); 4th/5th: adatom-free regions far away from/near (typically ~1–2 nm nearby) Fe adatoms ($\Omega$, 15/18.7 meV); 6th/7th: the same as 4th/5th but for Pb adatoms ($\Omega$, 14.5/12.5 meV).

, the doping-insensitive [6] and nonmagnetic phonon mode seems irrelevant. However, the experiment of well-controlled EPI suggests a collaborating effect of electron doping and EPI on the drastic $T_c$ enhancement [46], meaning the phonon effect is nonnegligible. Provided the contrast of phonon and spin fluctuation in magnetism, more detailed investigations are required to further distinguish them.

Figures 3(a) and 3(d) present $\Delta_1$ and $\Delta_2$ plotted versus the bosonic-mode energy $\Omega$, respectively. As the case of cuprates and iron pnictides [6–9,11], $\Delta_i$ is statistically anticorrelated with $\Omega$ (quantitatively analyzed in Fig. S8), demonstrating that the observed bosonic mode is an intrinsic collective excitation rather than a superconductivity-unrelated extrinsic inelastic excitation [47] or a band-structure effect [48]. The anticorrelation can be interpreted based on a local strong-coupling model [49]. As its application to FeSe/SiC [42], if electron-boson coupling (EBC) is independent of $\Delta_i$, $\Delta_i$ and $\Omega$ are then linked by

$$\Delta_i(r) = \Omega(r) e^{-\frac{\Omega(r)}{2N_0 g^2}}, \tag{1}$$

where $N_0$ is the DOS at Fermi level and $g$ is the spatially independent EBC constant. We quantified EBC strength by local normalized–tunneling-spectrum ratio $\sigma$ as Ref. [42], $\sigma = \frac{\mathrm{d}I/\mathrm{d}V(\Delta_2+\Omega)}{\mathrm{d}I/\mathrm{d}V(E_{\mathrm{dip}})}$, which shows no evident correlation with $\Delta_i$ [Figs. 3(c) and 3(f)]. Thus, according to Eq. (1), when $\Omega$ is comparatively larger, which is exactly the situation here, the exponential term dominates, resulting in the observed anticorrelated $\Delta_i$ and $\Omega$. Noteworthily, the seemingly $\Delta$-independent bosonic energy of 1–4-UC FeSe with and without K dosing in Ref. [13] appears contradicting the anticorrelation between $\Delta_i$ and $\Omega$ here. This is probably attributed to the mixing of $\Delta_1$ and $\Delta_2$ as roughly a single $\Delta$ and the relatively insufficient bosonic-energy data points for statistics in Ref. [13], tending to blur the underlying trend of the bosonic energy (Part VI in Ref. [41]).

Plotted in Figs. 3(b) and 3(e) are the ratios $\Omega/2\Delta_i$ vs. $\Delta_i$. Interestingly, $\Omega/2\Delta_i$ values are less than 1 in statistics [Fig. 3(b)] or all [Fig. 3(e)]. This characteristic was previously reported in spin-fluct-

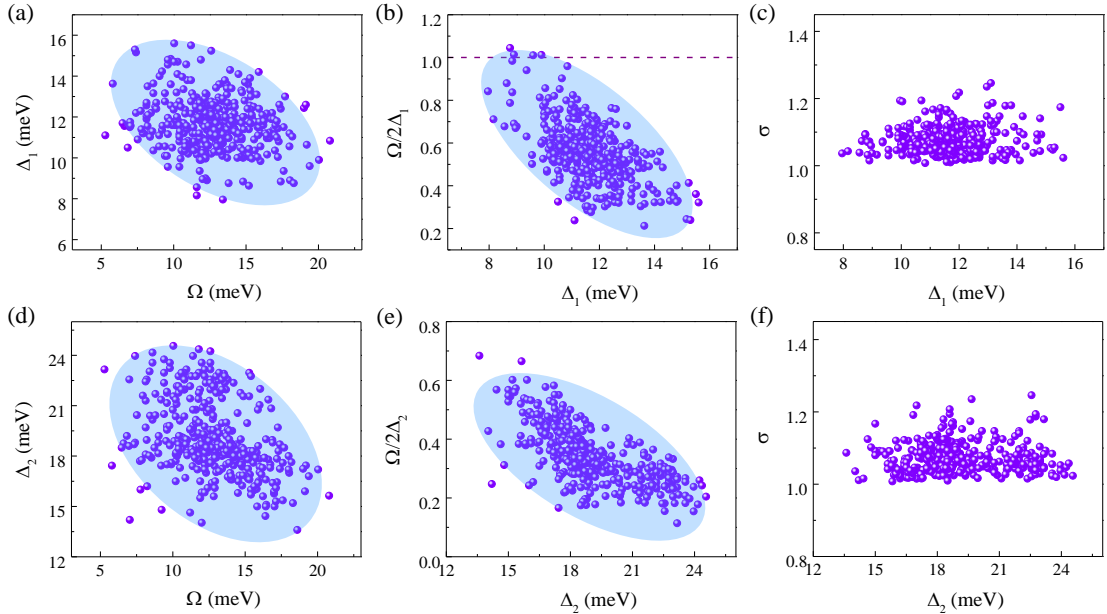

FIG. 3 (color online). (a) $\Delta_1$ plotted as a function of $\Omega$. (b) $\Omega/2\Delta_1$ and (c) $\sigma$ plotted as functions of $\Delta_1$. (d)–(f) The same as (a)–(c) but for $\Delta_2$ instead. $\Delta_i$, $\Omega$ and $\sigma$ are extracted from ~450 spectra.

uation–mediated copper-oxide and iron-based superconductors [5,7,11,42]. In Eliashberg theory of superconductors with a sign-reversing gap, the energy scale of spin resonance mode was predicted falling within 2Δ [50], consistent with our observations here. For another differently defined bosonic-mode energy, Ω' (Ω"), the anticorrelation between $\Delta_i$ and Ω' (Ω"), as well as $2\Delta_i$-restricted Ω' (Ω"), remains valid (Figs. S6 and S7). Combining these discussions, the magnetic-excitation origin of the bosonic mode is coincidentally favored in different degrees by all above methodologies, leaving the phonon interpretation likely irrelevant.

One question arises that whether the observed bosonic excitation plays a role of the pairing glue. To clarify such an issue, we imaged the local response of tunneling spectra to magnetically different Se defects. Normally, the imperfections appear in 1-UC FeSe film as multi-, single-Se vacancies, Se dimers and multimers (Fig. S2). The intrinsic defects are natural scatterers for Bogoliubov quasiparticles, whose scattering behaviors off magnetic and especially nonmagnetic scattering potentials have been well established to determine the paring symmetry of host superconductors [51]. Specifically, for conventional sign-preserving *s*-wave superconductors, Cooper pairs remain intact against potential scatterings, but are induced with in-gap quasiparticle bound states by magnetic scatterings [52]. In contrast, for sign-reversing multiband $s_\pm$-wave [53] and *d*-wave superconductors [54,55], both nonmagnetic and magnetic defects are pairbreakers that reconstruct the tunneling spectra by in-gap states.

Figure 4 presents the spatially resolved tunneling spectra approaching/across the multi-, single-Se vacancies [56] and the Se multimer. In detail, along a straight-line trajectory towards the multi-Se vacancy [Fig. 4 (a)], the tunneling spectra reveal in-gap bound states near −0.5 mV closely outside the vacancy edge [Fig. 4 (d)]. Meanwhile, the spectra imaged far enough away [e.g., bottom curves in Fig. 4(d)] exhibit nearly unchanged superconducting lineshapes. Apparently, the in-gap states are intimately correlated with the multi-Se vacancy. On the other hand, across the single-Se vacancy [Fig. 4(b)] and Se multimer [Fig. 4(c)], there emerge in the tunneling spectra both electron- and hole-like bound states within superconducting gaps [Figs. 4(e) and 4(f)]. The incomplete recovery of superconducting spectra, e.g., for the bottom and top curves in Figs. 4(e) and 4(f), is due to a shorter imaged distance (≤3 UC) away compared with the decay length of vacancy/multimer influence (~5 UC). When imaged at a defect-free region long enough distance away (≥5–10 UC), the tunneling spectra are recovered with a fully gapped multiband feature, ensuring the credibility of observed in-gap bound states.

Mainly in view of intrinsically distinguishing Se-dosage statuses [deficient (vacancy) vs. excessive (multimer)] and divergent responses of C4 symmetry of bound-state spatial distributions [broken (vacancy) vs. preserved (multimer)], the scattering potentials off multi-, single-Se vacancies and Se multimer are magnetically different (Part VII in Ref. [41]). As the bound states are induced by all three types of Se defects, especially nonmagnetic Se multimer (Part VII in Ref. [41]), the electron-phonon–coupled $s_{++}$-wave pairing is accordingly incompatible. Therefore, an unconventional sign-reversing pairing is revealed, associated with the magnetic excitation as the bosonic mode. Empirically and statistically in experiments, multiple bound states were induced by nonmagnetic scatterings in $s_\pm$-wave iron-based superconductors [53,57,58], in contrast to a single bound state by potential scatterings in *d*-wave cuprates [55,59]. Although we cannot absolutely exclude quasinodeless *d*-wave [33], the multipeak resonances induced by Se multimer phenomenologically prefer the extended $s_\pm$-wave pairing. Phase-sensitive quasiparticle-interference technique with an ultrahigh resolution will be helpful to more critically

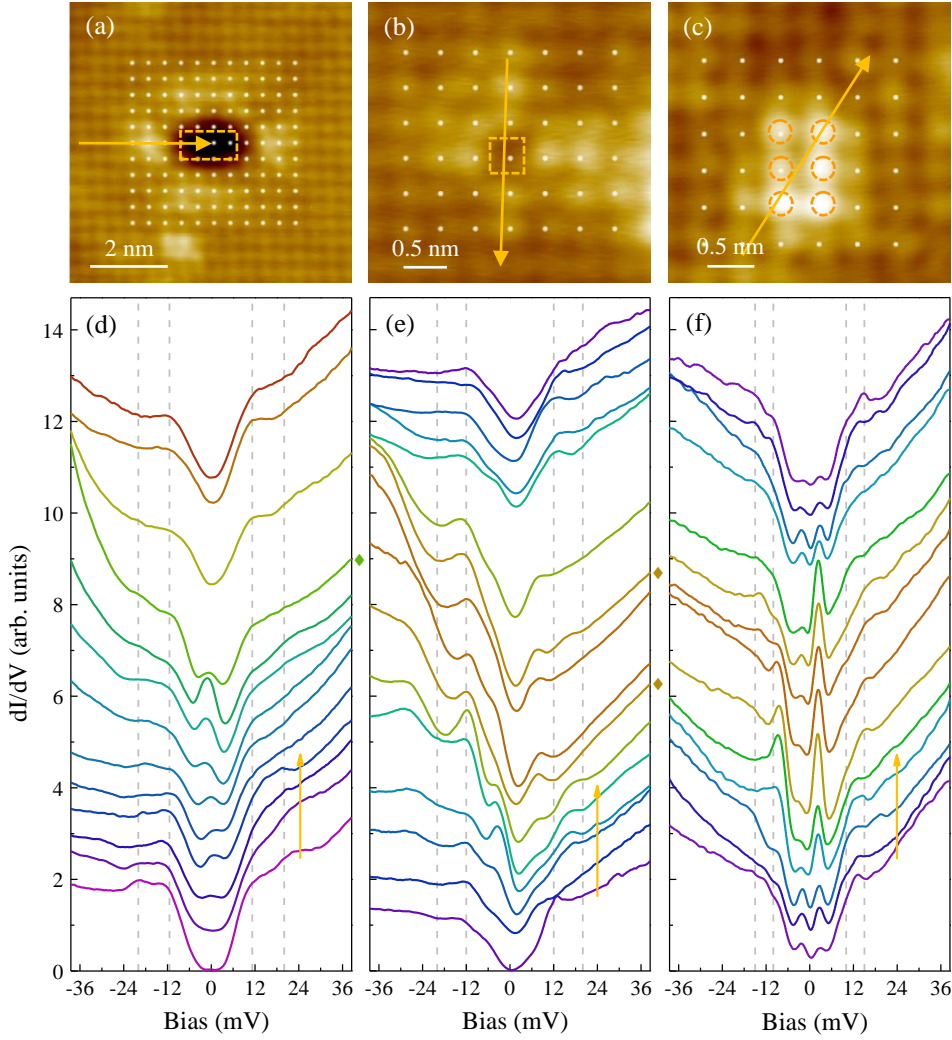

FIG. 4 (color online). (a)–(c) Topographic images of multi-, single-Se vacancies and Se multimer overlaid with surface Se lattice points (white dotted arrays). (d)–(f) Spatially resolved tunneling spectra along the arrows in (a)–(c), respectively, where the spectra imaged at vacancy edges are shown with rhombi in particular. The gray dashed lines are guides to the eyes of the bias positions of coherence peaks.

address the issue.

Finally, we discuss the physical nature of the detected bosonic excitation and its relation with the pairing glue. Originally, the phonon hump anomaly in conventional superconductors unambiguously establishes the phonon-mediated Cooper pairing in BCS theory [1–4]. Similarly, since the unconventional sign-reversing pairing attested by Se-defect scatterings generally yields a spin-fluctuation–mediated pairing component [51], we are encouraged to attribute the magnetic fluctuation to the bosonic hump. This reconciles our observations that, for instance, the bosonic mode appears anticorrelated with pairing strength and upper-bounded by twice the superconducting gap. Consequently, the magnetic excitation tends to be supported as the bosonic mode. With energy reasonably comparable with our bosonic-energy range, the aforementioned 21-meV spin resonance mode may be the concrete origin of the concluded magnetic excitation. It should be emphasized that the spin resonance was theoretically predicted in 1-UC FeSe [60] and experimentally detected in $(Li_{0.8}Fe_{0.2})OHFeSe$ [44], both at the vector connecting sign-reversing electron Fermi sheets at adjacent M points. These facts coincide with the scenario in extended $s_\pm$-wave (or quasinodeless $d$-wave) pairing, where antiferromagnetic spin fluctuation, with intensity peaked as a spin resonance, mediates the repulsive pair scatterings via intra- or predominantly inter-pocket vectors between phase-inversed Fermi surfaces as well. Therefore, the bosonic spin resonance is presumably the spin fluctuation that partially mediates the interfacial high-temperature superconductivity, which clarifies a nonnegligible role of electron–magnetic-excitation coupling in 1-UC FeSe.

The authors acknowledge helpful discussions with Dung-Hai Lee, Fan Zhang, Chen Chen and Yanzhao Liu. This work is financially supported by the National Basic Research Program of China (Grant No. 2018YFA0305600 and No. 2017YFA0303302), the National Natural Science Foundation of China (Grant No. 11774008) and the Key Research Program of the Chinese Academy of Sciences (Grant No. XDPB08-2).

[†]jianwangphysics@pku.edu.cn

# Supplemental Material

**I. Methods**

The Nb-doped STO(001) substrate was pretreated by the Se-flux method [1]. Then the 1-UC FeSe film was grown by coevaporating Fe (99.994%) and Se (99.999%) from an e-beam evaporator and a Knudsen cell, respectively, with the substrate held at 400 ℃, followed by annealing at 450 ℃ for hours. The Fe and Pb adatoms were dosed upon FeSe film at ~155–173 K. A polycrystalline PtIr tip was used throughout the experiments and the topographic images were obtained in a constant-current mode with a bias voltage applied to the tip. The scanning tunneling spectra were acquired using the standard lock-in technique with a typical bias modulation of 1 mV at 1.7759 kHz.

**II. Basic information of 1-UC FeSe film**

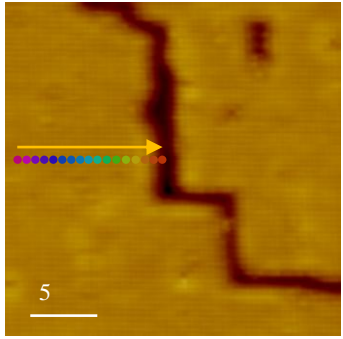

FIG. S1. Topographic image of a grain boundary, where the tunneling spectra in Fig. 2(a) are acquired along the same-colored dotted trajectory.

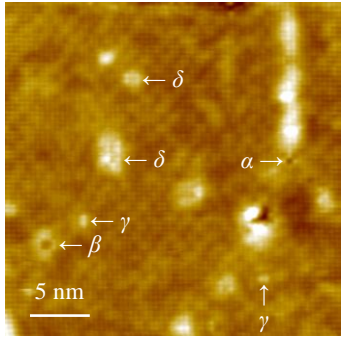

FIG. S2. Topographic image of another 1-UC FeSe film with randomly distributed intrinsic Se defects, e.g., single-($\alpha$), multi-Se vacancies ($\beta$) (dark spots), Se dimers ($\gamma$) and multimers ($\delta$) (bright spots), etc.

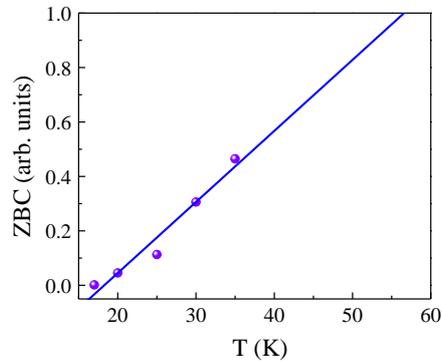

FIG. S3. Temperature-dependent zero-bias conductance (ZBC) extracted from Fig. 2(b), yielding an extrapolated critical temperature of 56.6 K at ZBC=1.

## III. Statistics of the other two differently defined hump energies

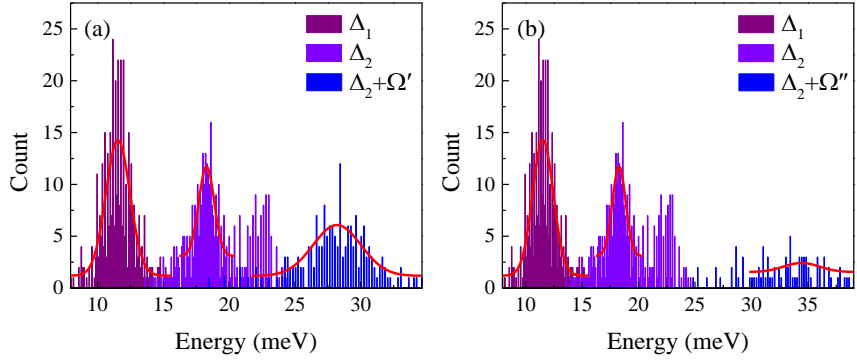

FIG. S4. Statistics of the double superconducting gaps, $\Delta_1$, $\Delta_2$, and the hump energy (a) $\Delta_2+\Omega'$ or (b) $\Delta_2+\Omega''$ for 1-UC FeSe film annealed at 450 ℃ for 3 h, where $\Delta_2+\Omega'$ and $\Delta_2+\Omega''$ are defined as the peak and dip energies (e.g., positive-bias side) of the derivative spectrum exemplified in Fig. 1(d). The Gaussian distributions (red) yield energy ranges of 6–13.7 and 12.2–20.1 meV for $\Omega'$ and $\Omega''$, respectively, which are also comparable with the 21-meV spin resonance mode [2] and the 20-meV optical phonon [3] only in terms of energy scales.

## IV. Bosonic-energy–determining details for the tunneling spectra in Fig. 2(d)

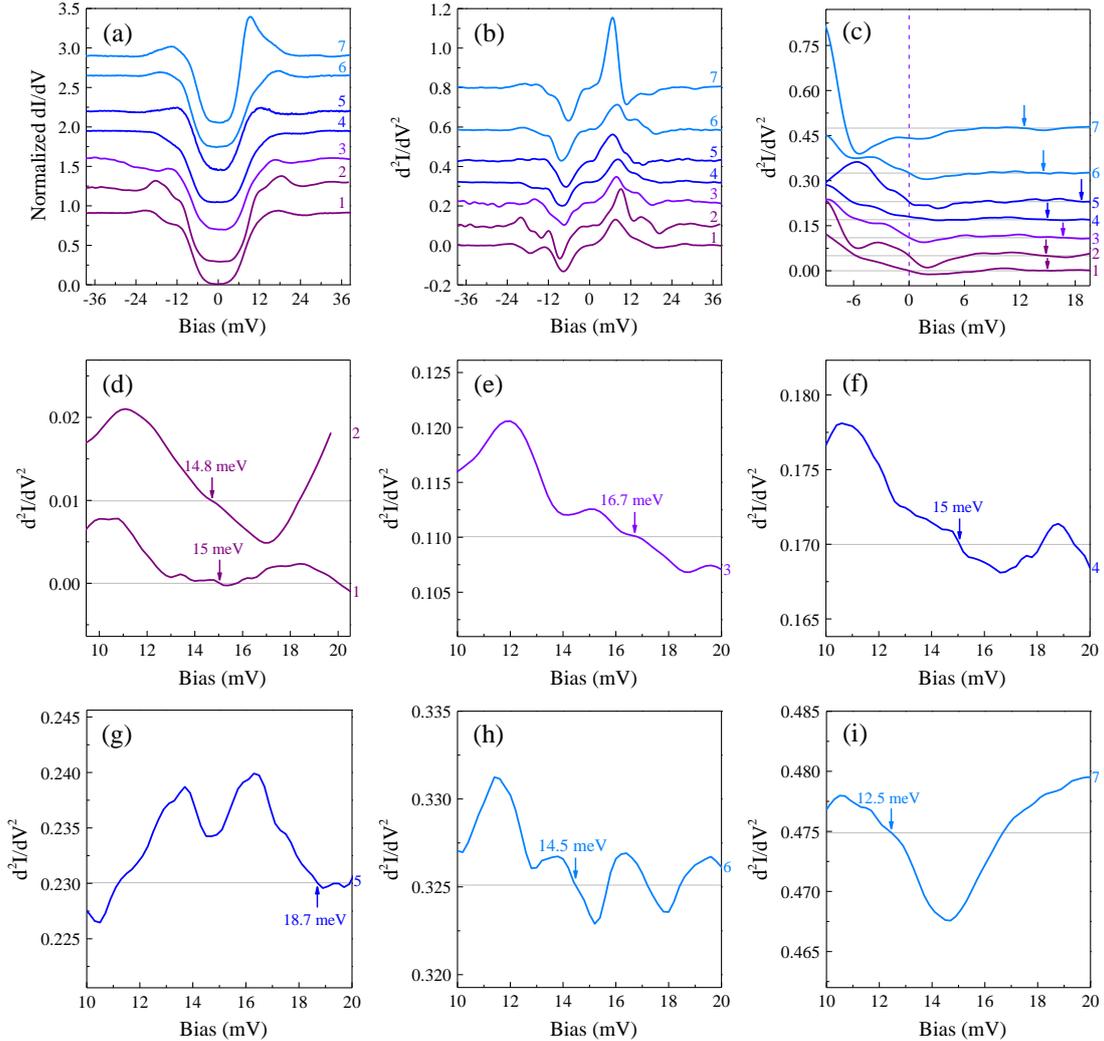

FIG. S5. (a) Averaged normalized tunneling spectra and (b)–(c) corresponding derivatives. All the spectra in (a)–(c) each are the same as those in Fig. 2(d) but are exhibited in different ways (in larger-bias windows and/or derivative forms). The gray horizontal lines in (c) mark the respective zero-$d^2I/dV^2$ lines before vertical offset. Note that, in (c), all the spectra have been shifted leftwards to center $\Delta_2$ at zero bias and only spectra ranging −9–19.6 mV are displayed to emphasize the zero points (arrows) of derivatives, which define the bosonic-mode energies $\Omega$ used in Fig. 2(d). (d–i) Close-up views of the 1st–7th derivative spectra in (c), highlighting the $\Omega$-determining (arrows) methodology in (c) and Fig. 2(d).

## V. Anticorrelation between pairing strength and bosonic energy

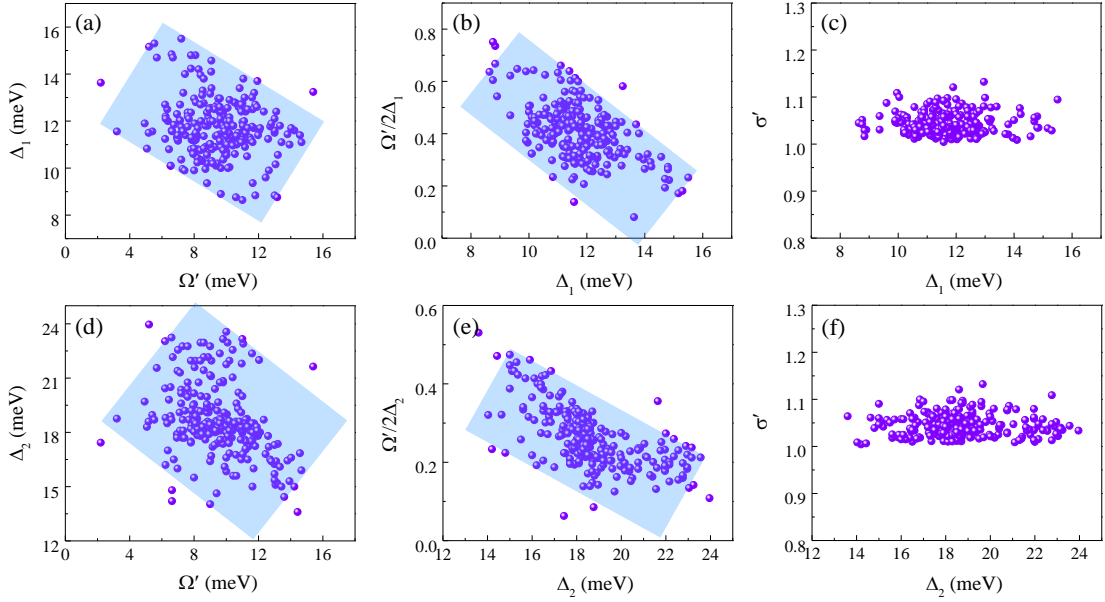

FIG. S6. The same as Fig. 3 but for a different bosonic-mode definition, $\Omega'$. $\sigma' = \frac{dI/dV(\Delta_2+\Omega')}{dI/dV(E_{\text{dip}})}$.

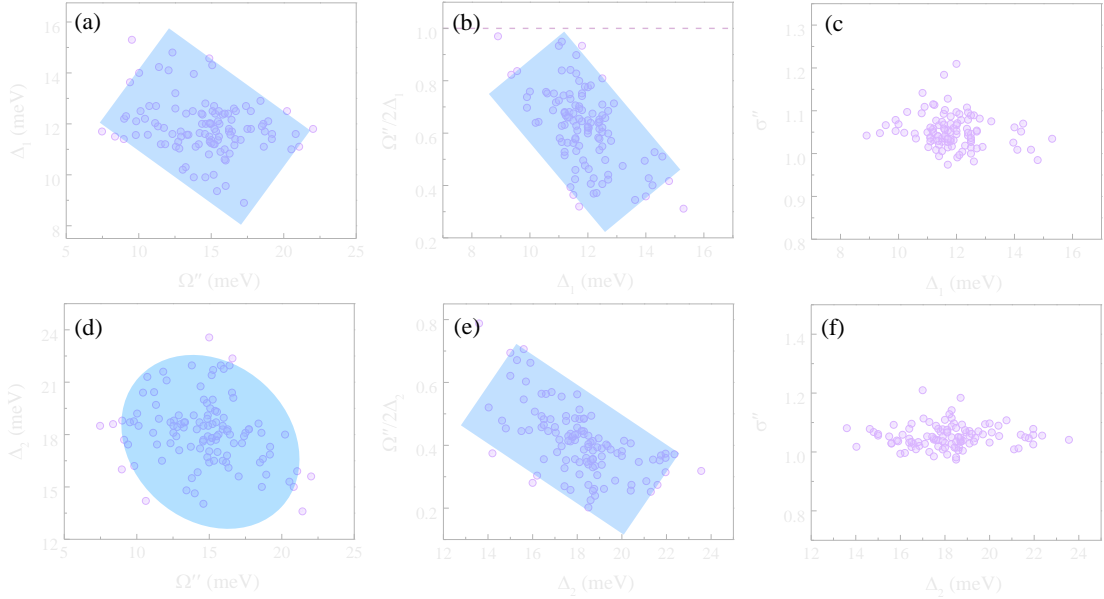

FIG. S7. The same as Fig. 3 but for a different bosonic-mode definition, $\Omega''$. $\sigma'' = \frac{dI/dV(\Delta_2+\Omega'')}{dI/dV(E_{\text{dip}})}$. It is seen that, although the bosonic-mode energy is defined in a slightly different manner as $\Omega'$ or $\Omega''$, the conclusions drawn in $\Omega$ case that anticorrelation between $\Delta_i$ and $\Omega$, $2\Delta_i$-restricted $\Omega$ and $\Delta_i$-independent $\sigma$ remain valid accordingly.

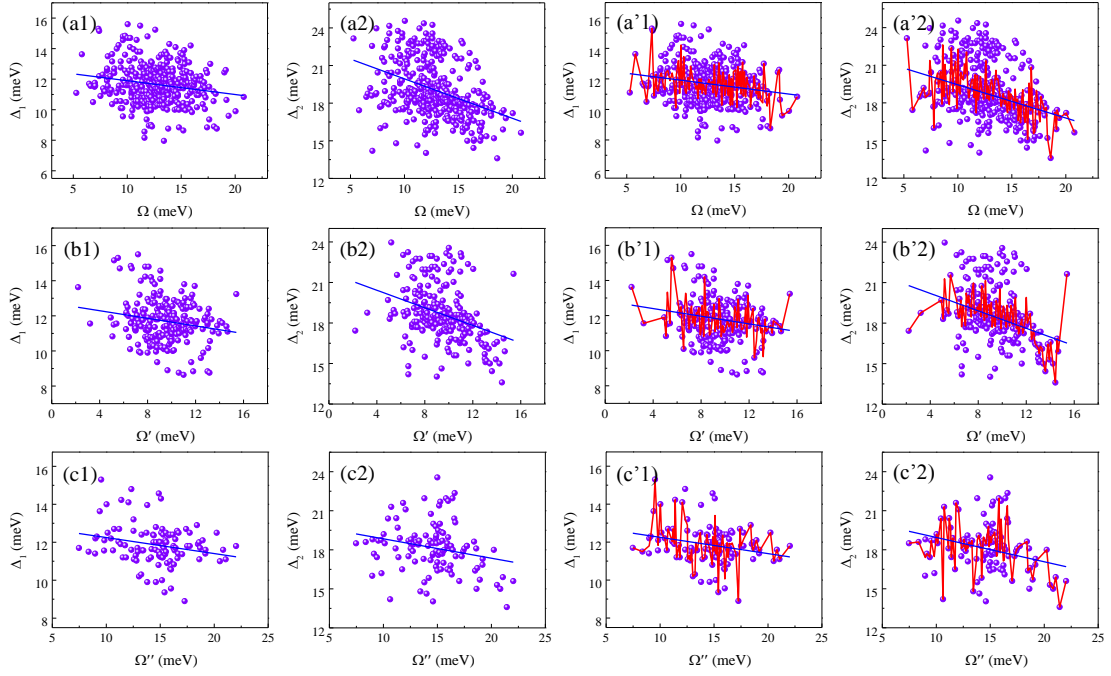

FIG. S8. Original (violet) and averaged $\Delta_i$ (red) plotted as functions of $\Omega$ ($\Omega'$, $\Omega''$), in which the latter is calculated from arithmetic mean of $\Delta_i$ corresponding to the same $\Omega$ ($\Omega'$, $\Omega''$) within 0.1-meV uncertainty. Anticorrelations between $\Delta_i$ and $\Omega$ ($\Omega'$, $\Omega''$) are determined by the negative slopes (a–c) in direct linear fittings (blue) of original $\Delta_i$ and (a'–c') in indirect linear fittings (blue) of averaged $\Delta_i$ as functions of $\Omega$ ($\Omega'$, $\Omega''$).

## VI. Comparison of Ref. [4] and our work for bosonic-energy vs superconducting-gap distributions

Noteworthily, the seemingly $\Delta$-independent bosonic energies $\Omega_1$ and $\Omega_2$ in Ref. [4] (Fig. S9) appear contradicting the anticorrelation between $\Delta_i$ and $\Omega$ in our FeSe film. Provided an ultrawide energy range of $\Delta$ (6.5–19 meV) and the failure in resolving double superconducting gaps, the $\Delta$ in Ref. [4] is probably a mixture of $\Delta_1$ and $\Delta_2$. This will make $\Omega_1$ and $\Omega_2$ disperse extensively along the horizontal $\Delta$ axis. Further considering the stretching of $\Delta$ axis along horizontal direction in the displayed figure and especially the relatively insufficient bosonic-energy data points for statistics, the truly existing $\Omega_1$-$\Delta$ and $\Omega_2$-$\Delta$ trends in Ref. [4] may be blurred and appear somewhat featureless (Fig. S9).

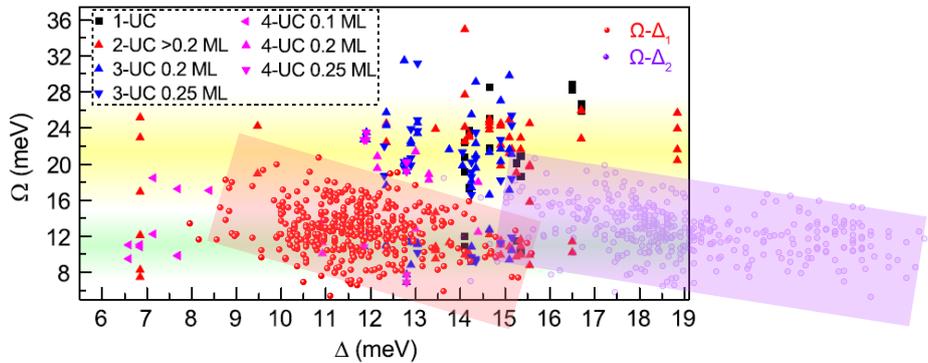

Fig. S9 The distribution of bosonic energy of 1–4-UC FeSe/STO with and without K dosing as a function of superconducting-gap magnitude $\Delta$ (adapted from Ref. [4]). $\Omega$ as functions of $\Delta_1$ (red) and $\Delta_2$

(violet) of our 1-UC FeSe film are overlaid for a direct comparison. ML, monolayer, defined as the coverage of K occupying all the hollow sites of surface Se lattice. For 1–4-UC FeSe in Ref. [4], the bosonic-energy distribution collapses into two groups, $\Omega_1$ and $\Omega_2$ ($\Omega_1<\Omega_2$). The $\Omega$ points in our FeSe film mainly match the $\Omega_1$ branch, with the $\Omega_2$ branch largely undetected, which may be due to the limited bias window of our tunneling spectra.

## VII. Magnetically different scattering potentials off multi-, single-Se vacancies and Se multimer

We present more spectrum data about intrinsic Se defects (mainly the spatial distributions of bound states), together with the discussions about their magnetic properties, in Part VII.

To visualize the defect-induced in-gap states [Figs. 4(d)–4(f)] in space, we imaged d$I$/d$V$ mappings near the biases of bound states for multi-, single-Se vacancies and Se multimer (Fig. S10). For the multi-Se vacancy, the resonance states display an anisotropic spatial distribution surrounding the vacancy [Fig. S10(a)]. The sharp contrast between regions outside and inside the vacancy indicates a localized property of the resonances near the vacancy edge. For the single-Se vacancy, while the electron-like bound states are comparatively more localized, the hole-like ones display a far larger spatial extent [Figs. S10(b) and S10(c)]. The spatial-distribution contrast of different resonances was rarely observed in previous reports [5,6], possibly revealing the resonance states as quasiparticle excitations with different origins. For the Se multimer, spatial distributions of the bound states reveal features [Figs. S10(d) and S10(e)] that intimately follow the multimer topography [Fig. 4(c)] somewhat with a slight offset yet.

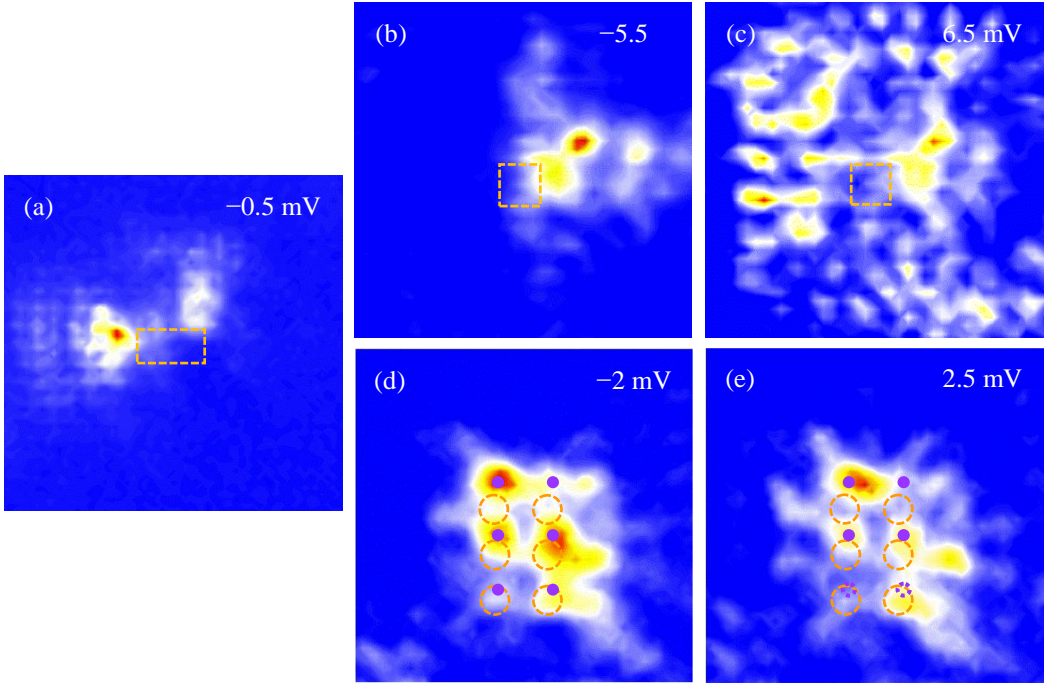

FIG. S10. d$I$/d$V$ mappings near the biases of bound states as labeled for (a) the multi-Se vacancy in Fig. 4(a), (b,c) the single-Se vacancy in Fig. 4(b) and (d,e) the Se multimer in Fig. 4(c), respectively. The violet dots in (d) and (e) are guides to the eyes of enhanced features in d$I$/d$V$ mappings.

Meanwhile, the tunneling spectra upon the Se dimer show enhanced (suppressed) spectral weights at electron (hole) states near the gap edge (Fig. S11), consistent with previous study [7]. The reconstructed spectral lineshapes also signal the presence of bound states as observed in Nb(110) [8], LiFeAs [9] and Na(Fe$_{0.96}$Co$_{0.03}$Cu$_{0.01}$)As [6]. As previously demonstrated, such type of bound states will be more clearly revealed if the fully gapped tunneling spectrum is subtracted from the abnormal one [6,8,9]. In essence, these "hidden" bound states probably have an origin in multiple angular-momentum channels in scatterings but are failed to be observed due to a limited spectrum resolution. With a possible identity of Fe vacancy [10], the nominal Se dimer cannot be fully regarded as a Se defect and will not be further discussed.

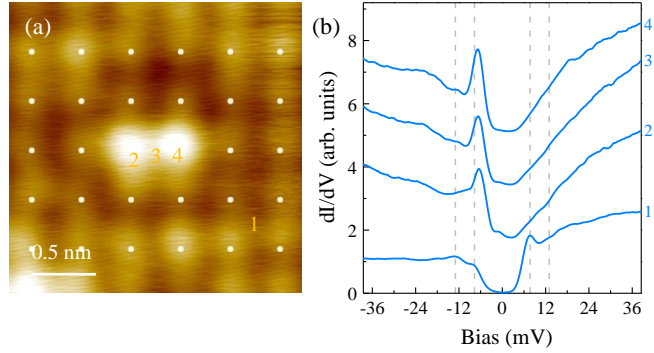

FIG. S11. (a) Topographic image of a Se dimer. (b) Tunneling spectra imaged at the positions as numbered in (a). The bound states evidenced by enhanced (suppressed) spectral weights at electron (hole) states near the gap edge are presented.

Identifying the magnetism "fingerprints" of impurities/defects in/on superconductors is crucial in concluding the pairing symmetry. By theoretical first-principles calculations [11] and experimental magnetic-exchange bias effect [12], the magnetic ground state of 1-UC FeSe/STO is demonstrated to be antiferromagnetic (AFM). Deduced from the chequerboard-like AFM spin configuration [11,12], the grain boundary along [100] or [010] direction of surface Se lattice (e.g., Fig. S1) is a ferromagnetic edge of mutually separated FeSe domains. Accordingly, the multi-Se vacancy that normally exposes the ferromagnetic edge of vacant FeSe segment [Fig. 4(a)] is presumably a magnetic defect. In view of intrinsically different Se-dosage statuses [deficient (vacancy) vs. excessive (multimer)] and nonmagnetism of Se atoms, the Se multimer is supposed to be magnetically different from the multi-Se vacancy and nonmagnetic in the limit of classical spin.

However, for the single-Se vacancy, judgement based on above chequerboard AFM scenario is inapplicable because of an ill-defined exposed "edge" [Fig. 4(b)]. Note that, in the nodal 30-UC FeSe/SiC, C2-symmetric resonance distributions in space were both observed near a magnetic vortex and a Fe adatom, as well as near the single-Se vacancy [13]. Since the vortex and Fe adatom are known as magnetic scatterers, the single-Se vacancy is possibly also magnetic. Analogously but in a slightly different manner, the short-range C4 symmetry of resonance distribution is broken near both multi- [Fig. S10(a)] and single-Se vacancies [Figs. S10(b) and S10(c)], while preserved near Se multimer in a unit of tetramer [Figs. S10(d) and S10(e)]. Basically, the resonance spatial mapping is modulated by superconducting coherence length $\xi(\boldsymbol{r})$ and reflects symmetry information of superconducting order parameter. Therefore, a preserved C4 symmetry of resonance distribution will be expected that originates from the gap-function symmetry of 1-UC FeSe/STO unless a magnetic pairbreaker is introduced. In this case, the multi-Se vacancy and Se multimer that breaks and preserves C4 symmetry, respectively, should be magnetic and nonmagnetic, respectively, coinciding with the judgement in the chequerboard AFM scenario. Taking above statements together, we are encouraged to regard the C4-breaking single-Se vacancy as a magnetic defect, consistent with the aforementioned deduction in the case of 30-UC FeSe/SiC. In conclusion, mainly in view of the different Se-dosage statuses and the divergent behaviors of C4 symmetry, the scattering potentials off multi-, single-Se vacancies and Se multimer are magnetically distinct.

Notice that, for the single-Se vacancy or Se multimer, the bound-state distributions in space are partially [Fig. S10(b)] or in a significant fraction [Fig. S10(c)–S10(e)] located upon the defect patterns. However, the quasizero-bias (near −0.5 mV) bound states induced by the magnetic multi-Se vacancy show a localized character preferring the vacancy edge [Fig. S10(a)]. Given the obviously different resonances of these Se defects in terms of near and deviating from zero bias, the vacancy-edge localization of distributed resonances displays a close correlation with quasizero bias particularly. Theoretically, the chiral edge modes with linear dispersions are predicted to produce quantized-conductance plateaus symmetric about zero bias, reminiscent of the spatially localized and energetically broadened quasizero-bias bound states induced by multi-Se vacancy. Thereby, further studies are required to deeply inspect whether topologically trivial or nontrivial quasiparticle excitations are the potential origins of these anomalous quasizero-bias states.